\documentclass[runningheads]{llncs}

\usepackage[T1]{fontenc}
\usepackage{graphicx}
\usepackage{amsmath}
\usepackage{amssymb}    %
\usepackage{multirow}
\usepackage{subcaption}
\usepackage{booktabs}
\usepackage{threeparttable}

\def\TPTminimum{\linewidth}

\makeatletter
\def\TPT@measurement{%
 \ifdim\wd\@tempboxb<\TPTminimum
  \hsize\TPTminimum
 \else
  \hsize\wd\@tempboxb
 \fi
 \xdef\TPT@hsize{\hsize\the\hsize \noexpand\@parboxrestore}\TPT@hsize
 \ifx\TPT@docapt\@undefined\else
  \TPT@docapt \vskip.2\baselineskip
 \fi \par \noindent\hfil\box\@tempboxb\hfil\par
 \ifvmode \prevdepth\z@ \fi
}
\makeatother

\usepackage{algorithm}
\usepackage{hyperref}

\usepackage{color}

\usepackage[n,operators,sets,keys,ff,mm,primitives]{cryptocode}
\AtBeginDocument{\pcfixhyperref}

\usepackage{xcolor}
\usepackage{tcolorbox}
\tcbuselibrary{breakable,skins}

\newtcolorbox{reviewblock}[1][Needs review]{
  enhanced,
  breakable,
  title={#1},
  fonttitle=\bfseries\small,
  colback=yellow!8,
  colbacktitle=orange!18,
  coltitle=black,
  frame hidden,
  borderline west={2pt}{0pt}{orange!70!black},
  arc=0pt,
  left=6pt,
  right=4pt,
  top=4pt,
  bottom=4pt,
  before skip=6pt,
  after skip=6pt
}

\newcommand{\sysname}{TESLA-for-5G}
\newcommand{\sysabbrv}{\textsf{TF5}}

\newcommand{\customsib}{\text{SIB-TF5}}
\newcommand{\blst}{\text{blst}}
\newcommand{\libsodium}{\text{libsodium}}

\newcommand{\coremark}{\textit{CoreMark}}
\newcommand{\ebacs}{\textit{eBACS}}
\newcommand{\schnorrhibs}{\textsf{Schnorr-HIBS}}
\newcommand{\etibs}{\textsf{E2IBS}}

\newcommand{\traceD}{\ensuremath{\mathcal{T}_D}}   %
\newcommand{\traceS}{\ensuremath{\mathcal{T}_S}}   %

\newcommand{\rrcconnected}{\texttt{RRC\_CONNECTED}}
\newcommand{\rrcidle}{\texttt{RRC\_IDLE}}
\newcommand{\rrcinactive}{\texttt{RRC\_INACTIVE}}

\newcommand{\mpk}{\ensuremath{\mathit{MPK}}}

\begin{document}

\hypersetup{hypertexnames=false}

\title{TESLA-for-5G: Broadcast Authentication for 5G Networks Using TESLA}

\author{Subin Song\inst{1} \and
  Michael K. Reiter\inst{2} \and
  Taekyoung (Ted) Kwon\inst{1}}
\authorrunning{S. Song et al.}
\institute{Seoul National University, Seoul, South Korea \\
  \email{\{sbsong66,tkkwon\}@snu.ac.kr} \and
  Duke University, Durham, NC, USA \\
  \email{michael.reiter@duke.edu}}

\maketitle

\begin{abstract}
5G base stations broadcast unauthenticated system information (SI) that every user equipment (UE) reads during cell selection. This enables attackers to broadcast forged SI from a fake base station~(FBS), deceiving UEs into camping on it.
Prior approaches to address this issue usually require UEs to authenticate System Information Block 1 (SIB1) using digital signatures. This necessitates computationally expensive verification for \emph{every} SIB1 reception, imposing a significant burden on resource-constrained UEs.
We propose \sysname{} (\sysabbrv{}), a broadcast authentication protocol for 5G SIB1 that combines TESLA with GG09 Schnorr-like identity-based signatures (IBS).
In the steady state, \sysabbrv{} enables UEs to authenticate each SIB1 message using a symmetric MAC and delayed key disclosure, eliminating the need for per-message digital signatures.
Initial trust is bootstrapped during cell entry using a lightweight GG09 IBS over the TESLA parameters, avoiding certificate distribution overhead.
We formally verify the security of \sysabbrv{} in Tamarin under a Dolev--Yao adversary
and demonstrate its favorable computation, communication, and storage costs through both an implementation on the
OpenAirInterface 5G stack and trace-driven analysis.
\end{abstract}

\section{Introduction}
\label{sec:introduction}

In 5G networks, base stations (gNBs) periodically broadcast unauthenticated system information, including the Master Information Block (MIB) and System Information Block Type~1 (SIB1). Every user equipment (UE) must acquire these blocks before camping on a cell.
By transmitting forged system information at a higher power, a fake base station~(FBS) can deceive nearby UEs into camping on it.
This enables attacks such as IMSI/SUCI catching, location tracking, protocol downgrades, and denial of service~\cite{hussainInsecureConnectionBootstrapping2019,tuckerDetectingIMSICatchersCharacterizing2025,chlosta5GSUCIcatchersStill2021}.
3GPP has explored countermeasures in TR~33.809~\cite{3gpp.33.809}, and prior research has proposed several solutions~\cite{hussainInsecureConnectionBootstrapping2019,singlaLookYouLeap2021,rossFixingInsecureCellular2024}. 
However, most prior proposals require digital signature verification for \emph{every} SIB1 reception.
What is worse, the UE not only needs to verify SIB1 on entering a new cell, but also on every re-acquisition of SIB1 when transitioning from \rrcconnected{} to \rrcidle{} without changing cells \cite{3gpp.38.331}.
These ``idle returns'' frequently occur over the day (see \S\ref{sec:eval:trace-cost}), which may impose a significant computational burden on resource-constrained UEs.

To address the problem, we present \sysname{} (\sysabbrv{}), an efficient broadcast authentication scheme for 5G SIB1 messages.
\sysabbrv{}~adopts the TESLA (Timed Efficient Stream Loss-tolerant Authentication) protocol~\cite{RFC4082-TESLA} to authenticate recurring SIB1 broadcasts using symmetric MACs.
UEs have to verify only one digital signature per cell for initial bootstrapping, and can then verify subsequent SIB1 messages with cheap MACs.
\sysabbrv{} uses a lightweight GG09 identity-based signature~\cite{GG09} to bootstrap TESLA commitment and parameters, eliminating the need for certificate distribution and validation.

Since TESLA uses a delayed key disclosure mechanism, it introduces an additional latency for verification. Some work raises concern about this delay~\cite{hussainInsecureConnectionBootstrapping2019}, but we show that TESLA's additional delay has minimal impact on user experience and thus acceptable for practical use (\S\ref{sec:design:motivation}).

Our contributions are as follows:
\begin{itemize}
  \item We design \sysabbrv{}, an authentication protocol that combines TESLA with GG09 IBS for 5G SIB1 broadcasts. We formally verify its security properties using the Tamarin Prover~\cite{meierTAMARINProverSymbolic2013} under a Dolev--Yao adversary model.
  \item We compare the computational, communication, and storage overheads of \sysabbrv{} with those of seven baseline schemes. We also implement \sysabbrv{} within the OpenAirInterface 5G stack~\cite{OpenAirInterface}, achieving an end-to-end camp completion latency of ${\approx}109$\,ms for returning UEs.
  \item We conduct a trace-driven analysis using real-world UE mobility and RRC state transition traces collected from two distinct environments. Our results show that \sysabbrv{}, with optimizations adapted to each environment, achieves a 55--65\% reduction in daily UE verification costs compared to a signature-only baseline.
\end{itemize}

\section{Background}
\label{sec:background}

\subsection{5G Broadcast and Fake Base Station Attacks}
\label{sec:background:fbs}

A 5G network consists of User Equipment (UE), Next-Generation RAN (NG-RAN) base stations (gNBs), and the 5G Core (5GC)~\cite{3gpp.23.501}.
Each gNB periodically broadcasts Master Information Block (MIB) and System Information Block Type~1 (SIB1) messages containing the essential parameters UEs require for network access~\cite{3gpp.38.300}.
Crucially, these broadcasts lack authentication: UEs cannot verify whether a received MIB or SIB1 originates from a legitimate gNB~\cite{hussainInsecureConnectionBootstrapping2019}.
Consequently, an adversary can deploy a rogue gNB that broadcasts forged system information at a higher power, tricking nearby UEs into camping on a fake cell.
While 5G provides mutual authentication via 5G-AKA, this activates only \emph{after} cell selection, leaving the initial phase vulnerable~\cite{gaoEvaluatingDelegatedDigital2021,purificationFakeBaseStation2024}.
Although 3GPP has explored countermeasures such as identity-based signatures in TR~33.809~\cite{3gpp.33.809}, no broadcast authentication mechanism has been standardized to date~\cite{3gpp.33.809,singlaLookYouLeap2021}.

\subsection{TESLA Broadcast Authentication Protocol}
\label{sec:background:tesla}

TESLA (Timed Efficient Stream Loss-tolerant Authentication)~\cite{RFC4082-TESLA} uses delayed key disclosure to authenticate broadcast messages with symmetric primitives: MACs and one-way hash functions.
Since anyone who knows the MAC key can forge messages, TESLA keeps the key secret during use and reveals it later.
The receiver buffers the received packet until the key is disclosed; once they receive the key after some delay, they can use it to verify the previously buffered message.

On protocol startup, the sender repeatedly applies a one-way function $F$ to a random seed $K_N$ to construct the key chain:
\[
K_N \xrightarrow{F} K_{N-1} \xrightarrow{F} \cdots \xrightarrow{F} K_1 \xrightarrow{F} K_0,
\]
$K_i$ denotes the key for interval $i$, and $K_0$ is the commitment to the chain.
Therefore, the keys are used in reverse order of their generation ($K_1, K_2, \ldots$).
If an adversary knows $K_i$, they can compute $K_{i-1}, \ldots, K_0$ but not $K_{i+1}, \ldots, K_N$.
That is, if the adversary knows the key for interval $i$, they can forge messages for intervals $0$ through $i$, but not for any future intervals.

To start using TESLA, the receiver must first obtain the commitment $K_0$ and the TESLA parameters (interval duration, disclosure delay, etc.) through an authenticated channel, e.g., a digital signature.
TESLA parameters include $T_0$ (the start time of interval~0), $T_{\mathrm{int}}$ (the interval duration), $d$ (key disclosure delay in intervals), and the chain length~$N$.
For TESLA to be secure, the receiver's clock must be loosely synchronized with the sender's clock, and the receiver must know an upper bound $D_t$ on the sender's clock drift relative to its own clock.
Starting at $T_0$, time is divided into $N$ intervals of duration $T_{\mathrm{int}}$.
During interval $i$, the sender computes MACs with $K_i'=F'(K_i)$ and discloses $K_{i-d}$, so $K_i$ remains secret until interval $i+d$.

On receiving a packet sent in interval $i$, the receiver computes an upper bound on the sender's current interval using the packet reception time and $D_t$ (see Appendix~\ref{app:safe-packet-test} for details) and buffers the packet only if the sender cannot yet have reached $i+d$.
This safe packet test ensures that the MAC key is still secret at reception, so that the packet was not forged by an adversary who has already learned the key.
When $K_i$ arrives after $d$ intervals, the receiver verifies the disclosed key by checking $F^i(K_i) = K_0$, where $F^n(i) = F(F^{n-1}(i))$ with $F^0(i)=i$. Then, the receiver verifies the buffered MAC with $K_i'=F'(K_i)$.

\subsection{Identity-Based Signature}
\label{sec:background:ibs}

Shamir~\cite{shamirIdentityBasedCryptosystemsSignature1985} introduced
\emph{identity-based cryptography}, which derives a user's public key directly from an identity string, eliminating the need for certificates.
In an identity-based signature (IBS) scheme, a trusted authority called the
\emph{Private Key Generator} (PKG) holds a master secret key ($\mathit{msk}$)
with the corresponding master public key ($\mathit{MPK}$) and issues each
user a signing key for their identity.
Anyone can verify a signature with only $\mathit{MPK}$ and the signer's identity.
Because no certificate needs to be transmitted with the signature, IBS is well suited to environments with strict message size
limits, such as cellular broadcast.
3GPP TR~33.809 includes IBS as a candidate authentication scheme for 5G broadcast messages~\cite{3gpp.33.809}.

The standard security notion for IBS is \emph{existential unforgeability
under adaptive chosen-message and chosen-identity attack}
(EUF-IBS-CMA)~\cite{bellareSecurityProofsIdentityBased2004,GG09}.\footnote{The
exact name varies; we follow the terminology of~\cite{GG09}.}
An adversary may adaptively query an extraction oracle for signing keys of
chosen identities and a signing oracle for arbitrary identity--message pairs.
It wins by forging a valid signature on an identity--message pair for
which it neither extracted the key nor obtained a signature.

Many existing IBS constructions are pairing-based~\cite{HessIBS,BLMQ}, which rely on bilinear maps and are often expensive to compute. In contrast, pairing-free IBS schemes, such as GG09~\cite{GG09} avoid such expensive operations, making them efficient for resource-constrained devices. We adopt the GG09 scheme in our work, which is described in detail in the following section.

\subsubsection{GG09 IBS Scheme}
\label{sec:background:gg09}

Galindo and Garcia~\cite{GG09} construct a pairing-free IBS from two
concatenated Schnorr signatures on standard elliptic curves.
The PKG uses the first to derive a signing key from the identity, and the
signer uses the second to sign a message.
Let $\mathbb{G}$ be a cyclic group of prime order~$q$ with generator~$P$,
$H_1, H_2\colon \{0,1\}^* \to \mathbb{Z}_q$ be hash functions modeled as
random oracles, and $\mathit{id}$ be the signer's identity.
GG09 IBS consists of four algorithms:
\begin{itemize}
  \item $\mathsf{Setup}() \to (\mathit{MPK}, \mathit{msk})$: The PKG chooses
    $\mathit{msk} \xleftarrow{} \mathbb{Z}_q$ and sets
    $\mathit{MPK} = \mathit{msk}P$.
  \item $\mathsf{Extract}(\mathit{msk}, \mathit{id}) \to \mathit{sk}_{\mathit{id}}$:
    The PKG chooses
    $r \xleftarrow{} \mathbb{Z}_q$ and computes $R = rP$,
    $c = H_1(R,\, \mathit{id})$, and
    $y = r + \mathit{msk} \cdot c \bmod q$, yielding
    $\mathit{sk}_{\mathit{id}} = (y,\, R)$.
  \item $\mathsf{Sign}(\mathit{sk}_{\mathit{id}}, \mathit{id}, m) \to \sigma$:
    The signer parses
    $\mathit{sk}_{\mathit{id}} = (y, R)$, chooses $a \xleftarrow{} \mathbb{Z}_q$,
    and computes $A = aP$, $e = H_2(\mathit{id},\, A,\, m)$, and
    $s = a + y \cdot e \bmod q$, yielding $\sigma = (A,\, s,\, R)$.
  \item $\mathsf{Verify}(\mathit{MPK}, \mathit{id}, m, \sigma) \to \{\top,\bot\}$:
    The verifier parses
    $\sigma = (A, s, R)$, computes $c = H_1(R, \mathit{id})$ and
    $e = H_2(\mathit{id}, A, m)$, and accepts iff
    $sP = A + e(R + c \cdot \mathit{MPK})$.
\end{itemize}

Correctness follows from the extraction relation $yP = R + c \cdot \mathit{MPK}$
and the signing relation
$sP = A + e \cdot yP = A + e(R + c \cdot \mathit{MPK})$.
The authors' original EUF-IBS-CMA security proof under the discrete logarithm assumption in the
random oracle model~\cite{GG09} was found to contain flaws in both
reductions~\cite{chatterjeeGalindoGarciaIdentityBasedSignature2013};
Chatterjee et al.~\cite{chatterjeeGalindoGarciaIdentityBasedSignature2013} provided a corrected proof that re-establishes security
under the same assumption with a non-tight
reduction.
The scheme itself remains unbroken, though a tight reduction remains open.

\section{Assumptions and Security Goals}
\label{sec:threat-model}

\subsubsection{Adversary model.}
\label{sec:threat-model:adversary}
We assume a Dolev--Yao adversary consistent with the TESLA threat model~\cite{RFC4082-TESLA} and prior SI authentication work~\cite{hussainInsecureConnectionBootstrapping2019,singlaLookYouLeap2021,rossFixingInsecureCellular2024}.
The adversary controls a localized wireless channel and can eavesdrop on, inject, modify, drop, or delay any over-the-air packet.
It can also deploy rogue gNBs that broadcast arbitrary system information at arbitrary signal strength.
A legitimate gNB's IBS signing key may also be compromised, since equipment at a cell site is physically exposed.
The adversary is computationally bounded and cannot break standard cryptographic assumptions, such as the discrete logarithm problem and the security of the underlying hash functions.

We exclude PKG compromise, relay attacks (see \S\ref{sec:discussion:relay}), side-channel attacks, and denial of service caused by jamming.

\subsubsection{PKG deployment.}

No authoritative document specifies or recommends where or how to deploy the PKG when using IBS for authenticating SI broadcasts. 3GPP TR 33.809~\cite{3gpp.33.809} also leaves this unspecified.
Existing PKI certificates used by gNBs for mutual authentication are managed by the operator's PKI CA and RA~\cite{3gpp.33.310}, but 3GPP does not define which NF hosts the CA/RA or whether it is operated as an independent service.
Following this precedent, we do not assign the PKG to a specific NF, but allow the operator to choose to operate it inside or outside the core network.
Meanwhile, as mentioned above, we exclude PKG compromise from our threat model. We also assume that gNBs communicate with the PKG over a secure and authenticated channel, just like the operator's PKI CA/RA and other NFs in the core network.

\subsubsection{Assumptions.}
\label{sec:threat-model:assumptions}

The network operator runs a PKG that is not compromised, and the PKG's master public key (\textit{MPK}) is pre-provisioned to UEs via a secure channel such as SIM provisioning.
Through a secure channel, the PKG issues IBS signing keys to gNBs, bound to their cell IDs and key validity periods.

We assume that gNBs periodically synchronize their own clock using GPS, and ensure that they keep their clock error within a fixed bound from GPS time~\cite{lombardiEvaluationDependenciesCritical2021,itu.G8271.2020}.
We also assume that UEs have a GPS receiver and can synchronize their own clock to GPS time on demand.
Time synchronization is discussed in more detail in \S\ref{sec:design:timesync}.

\subsubsection{Security goals.}
\label{sec:threat-model:goals}
The primary goal of \sysabbrv{} is to enable UEs to reject SIB1 messages not originating from a legitimate gNB, thereby \textbf{preventing camping on fake base stations}.
Specifically, \sysabbrv{} provides the following properties \emph{when the gNB's signing key is not compromised}:
\begin{itemize}
  \item \emph{Source authentication.} The UE only accepts SIB1 messages if it originates from a legitimate gNB.
  \item \emph{Message integrity.} The UE can detect and reject SIB1 messages that have been modified in transit.
  \item \emph{Replay protection.} The UE can detect and reject replayed SIB1 messages from previous TESLA intervals.
\end{itemize}

\sysabbrv{} does \emph{not} provide non-repudiation or pre-authentication DoS protection. 

\section{Protocol Design}
\label{sec:design}

\subsection{Motivation}
\label{sec:design:motivation}

Many previous SI authentication work authenticate SIBs using digital signatures~\cite{hussainInsecureConnectionBootstrapping2019,singlaLookYouLeap2021,rossFixingInsecureCellular2024}. These approaches require the UE to verify a signature for every SIB acquisition.
The problem is that a UE does not only need to acquire SIB1 once per cell; it repeatedly has to reacquire it while staying in the cell, when it transitions from \rrcconnected{} to \rrcidle{}.
It cannot even use a cached copy, as mandated by 3GPP 38.331~\cite{3gpp.38.331}.
The TESLA protocol~\cite{RFC4082-TESLA} can reduce these recurring verification costs by authenticating only the initial chain commitment with a signature and verifying subsequent SIB broadcasts using symmetric MACs.
However, Hussain et al.~\cite{hussainInsecureConnectionBootstrapping2019} deemed TESLA infeasible for SI authentication because TESLA has an inherent key disclosure delay, which increases authentication latency.

Still, we observe that these frequent ``recurring'' SIB1 acquisitions are often followed by an idle period, which hides the authentication delay from the user.
When the UE has to acquire SIB1 due to transitioning from \rrcconnected{} to \rrcidle{} (we call this transition an ``idle return''), the UE typically stays idle for a while after the transition. Therefore, it is likely that it will have enough time to complete authentication before it needs an actual connection.

Analysis of Meng et al.~\cite{mengModelingGeneratingControlPlane2023} supports that the UE's idle state is often long enough to accommodate the authentication delay. 
They analyzed one week of MME signaling logs from $23{,}388$ phones on a commercial LTE network in the United States in 2022, and sampled a cluster of phones to study the distribution of idle durations (i.e., the time for which an UE stays in \rrcidle{} state). Figure~4(b) in their paper shows that roughly $95\%$ or more of the idle durations exceed $320$\,ms, and $90\%$ or more exceed $1$\,s. 
They also report that they ``observe similar results for all the other clusters''. This suggests that even the worst case delay of \sysabbrv{} (i.e., $320$\,ms\footnote{\emph{Worst case} delay $320$\,ms: 160\,ms for key disclosure~(\S\ref{sec:design:timesync}) and 160\,ms for bootstrap~(\S\ref{sec:eval:latency}). The \emph{average} delay is shorter because the UE likely joins in the middle of a TESLA interval (see Table~\ref{tab:eval-latency}).}) is likely to be hidden by the idle period in most cases, and thus has \emph{minimal impact} on user experience.

UEs also have to reacquire SIB1 on \emph{cell reselection}~\cite{3gpp.38.331}, which is a frequent event as well. Thankfully, cell reselection happens only during \rrcidle{} or \rrcinactive{}~\cite{3gpp.38.304}, so the same rationale applies: the UE likely has enough time for authentication. Handovers during active connection are handled by the network and UEs do not have to read broadcast SIB1~\cite{3gpp.38.304,3gpp.38.331}.

The only cases that TESLA delays matter are initial registration (e.g., device power-on), returning from out of service, and radio link failure (i.e., the UE loses connection to the gNB while \rrcconnected{} and has to reselect a new cell).
In Appendix~\ref{app:delay-sensitive-event-frequency}, we summarize prior work that quantifies the frequency of these events, and show that they are infrequent.
SI modification and PWS alerts also happen infrequently and are unrelated to UE connectivity, so are unlikely to affect user experience. 

Based on these observations, we conclude that TESLA is practical for SI authentication, and we design \sysabbrv{} to integrate TESLA with an IBS scheme to efficiently authenticate SIB1 broadcasts.

\subsection{Overview}

\sysabbrv{} integrates TESLA~\cite{RFC4082-TESLA} with the GG09~\cite{GG09} IBS scheme.
In this design, IBS is used to bootstrap TESLA root commitment and parameters, while TESLA authenticates each SIB1 broadcast through MACs.
By deriving the gNB's public key from its Cell ID and key validity period, IBS eliminates the overhead of certificate distribution and validation.
\sysabbrv{} involves three entities: the PKG, gNB, and UE.
The \emph{PKG}, run by the network operator, holds the master key pair $(\mathit{MPK}, \mathit{msk})$ and issues IBS signing keys to gNBs.
At startup, the \emph{gNB} generates a TESLA key chain and periodically broadcasts the chain commitment and parameters in a dedicated \customsib{} signed with its IBS signing key.
Subsequent SIB1 broadcasts include a MAC tag and disclosed TESLA key.
Upon entering a cell, a \emph{UE} verifies the IBS signature in \customsib{} to obtain the chain commitment, and then verifies subsequent SIB1 messages using TESLA MAC.

While \sysabbrv{} authenticates SIB1, it does not authenticate the MIB since the PBCH carrying the MIB has a fixed 32-bit payload~\cite{3gpp.38.212} that cannot accommodate an authentication tag.
An unauthenticated MIB only exposes the UE to denial of service (DoS) attacks \cite{hussainInsecureConnectionBootstrapping2019}.
In contrast, more critical attacks---such as IMSI/SUCI catching~\cite{chlosta5GSUCIcatchersStill2021}, location tracking, downgrade, and power draining~\cite{mubasshirGottaDetectemAll2025}---require the UE to read SIB1 and camp on the fake base station.
Our authentication scope aligns with prior work: \cite{hussainInsecureConnectionBootstrapping2019,singlaLookYouLeap2021,rossFixingInsecureCellular2024} all authenticate only SIBs, excluding the MIB.

\subsection{Protocol Description}
\label{sec:design:protocol}

\subsubsection{Notation.}
Let $N$ denote the TESLA chain length and $\widehat{K}_{i-d}$ a candidate value for the disclosed key that has not yet been verified, where $i$ is the current TESLA interval index and $d$ is the key disclosure delay.
We define the iterated hash function as $F^n(x) = F(F^{n-1}(x))$ with $F^0(x)=x$, and let $(\mathit{MPK}, \mathit{msk})$ be the PKG's master key pair.
$W$ denotes the UE's signature freshness window, and the TESLA parameter set is denoted by $\Pi = (T_0, T_{\mathrm{int}}, d, N, K_0)$.
$\epsilon_{\mathit{g}}^\mathrm{max}$ is the maximum timing error allowed for the gNB at any time (fixed, public parameter); $\epsilon_{\mathit{u}}(t)$ denotes the UE's calculated clock error bound at time $t$~(see \S\ref{sec:design:timesync} for more details on timing).
Symbols for per-message fields are summarized in Table~\ref{tab:wire-format}.

\subsubsection{Protocol.}
\begin{table}[tbp]
\caption{\sysabbrv{} wire format. \textbf{Size}, \textbf{Auth}, and \textbf{Symb.} denote each field's size, whether the field is authenticated (by TESLA MAC for the SIB1 extension, by GG09 signature for \customsib{}), and its symbolic notation, respectively.
}
\label{tab:wire-format}
    \begin{subtable}{0.45\textwidth}
        \centering
        \captionsetup{justification=raggedright,singlelinecheck=false}
        \caption{SIB1 extension for \sysabbrv{}.}
        \label{tab:sib1-extension-format}
        \begin{threeparttable}
        \begin{tabular}{lrcl}
        \hline
        \textbf{Field} & \textbf{Size} & \textbf{Auth} & \textbf{Symb.} \\
        \hline
        $\mathit{sp}^\mathrm{next}$ flag\tnote{*} & 1B & $\checkmark$ & $\mathit{sp}^\mathrm{next}$ \\
        \hline
        Interval index & 4B & $\times$ & $i$ \\
        Disclosed key & 16B & $\times$ & $\widehat K_{i-d}$ \\
        Next-chain comm. & 16B & $\checkmark$ & $K_0^{\mathrm{next}}$ \\
        TESLA MAC tag  & 16B & --- & $\tau$ \\
        \hline
        Total & 53B & & \\
        \hline
        \end{tabular}
        \begin{tablenotes}\footnotesize
            \item The original SIB1 content is not shown in this table, but it is authenticated by MAC.
            \item[*] See \S\ref{sec:design:chain-renewal}.
        \end{tablenotes}
        \end{threeparttable}
    \end{subtable}
    \hfill
    \begin{subtable}{0.54\textwidth}
        \centering
        \captionsetup{justification=raggedright,singlelinecheck=false}
        \caption{\customsib{} bootstrap message.}
        \label{tab:sibtf5-extension-format}
        \begin{tabular}{lrcl}
        \hline
        \textbf{Field} & \textbf{Size} & \textbf{Auth} & \textbf{Symb.} \\
        \hline
        Cell ID & 5B & $\checkmark$ & $\mathit{cid}$ \\
        \hline
        Starting time & 3B & $\checkmark$ & $T_0$ \\
        Interval duration (ms) & 3B & $\checkmark$ & $T_{\mathrm{int}}$ \\
        Key disclosure delay & 1B & $\checkmark$ & $d$ \\
        Chain length & 4B & $\checkmark$ & $N$ \\
        Chain commitment & 16B & $\checkmark$ & $K_0$ \\
        \hline
        Signing key expiration & 3B & $\checkmark$ & $t_{\mathrm{exp}}$ \\
        GG09 signature & 96B & --- & $\sigma$ \\
        Signing timestamp & 3B & $\checkmark$  & $t_{\mathrm{sign}}$ \\
        \hline
        Total & 134B & &\\
        \hline
        \end{tabular}
    \end{subtable}
\end{table}

\sysabbrv{} operates in two phases: the bootstrap phase and the TESLA phase.
In the \emph{bootstrap phase}, the UE verifies a GG09 IBS signature on a dedicated \customsib{} message to obtain a trusted TESLA chain commitment and parameters.
In the subsequent \emph{TESLA phase}, the UE verifies each SIB1 broadcast using its TESLA MAC tag whose key is disclosed in a later interval.
Table~\ref{tab:wire-format} shows the wire formats for \sysabbrv{} messages.
We use a custom SIB, \customsib{}, to transmit the TESLA parameters and their GG09 signature, while the TESLA MAC and disclosed key are transmitted along with the original SIB1 content in a \texttt{lateNonCriticalExtension} field~\cite{3gpp.38.331}.
We use the custom SIB for bootstrapping to avoid the bandwidth cost of transmitting a full signature on every SIB1 repetition.

\begin{algorithm}[tbp]
  \caption{\sysabbrv{} UE-side authentication procedures.}
  \label{alg:ue-procedures}
  \raggedright
  \procedure[linenumbering, skipfirstln, codesize=\small]{$\mathsf{TF5.Bootstrap}(P,\; t_{\mathrm{rx}},\; \epsilon_{\mathit{u}}^{\mathrm{rx}},\; \epsilon_{\mathit{g}}^\mathrm{max},\; \mathit{MPK},\; W)$}{%
    \textbf{Input:}\; P \text{ : received \customsib{} packet};\; t_{\mathrm{rx}} \text{ : reception time};\; \epsilon_{\mathit{g}}^\mathrm{max};\; \mathit{MPK};\; W \pcskipln \\
    \epsilon_{\mathit{u}}^{\mathrm{rx}} = \epsilon_{\mathit{u}}(t_{\mathrm{rx}}) \text{ : UE clock error bound at } t_{\mathrm{rx}} \text{; see \S\ref{sec:design:timesync} and Appendix~\ref{app:clock-error-bound}} \pcskipln \\
    \textbf{Output:}\; \top \text{ if bootstrap succeeds, } \bot \text{ otherwise} \pcskipln \\
    \textbf{State:}\; \Pi,\; \mathit{cid},\; K_{\mathrm{anc}},\; i_{\mathrm{anc}},\; \mathit{buf},\; K_0^{\mathrm{next}},\; \mathit{sp}^\mathrm{next} \\
    (\widehat{\mathit{cid}},\, \widehat{\Pi},\, \sigma,\, t_{\mathrm{sign}},\, t_{\mathrm{exp}}) \gets P \pccomment{$\sigma$: signature} \\
    \pcif t_{\mathrm{rx}}+\epsilon_{\mathit{u}}^{\mathrm{rx}} > t_{\mathrm{exp}} \pcthen \pcreturn \bot \pccomment{gNB key expired} \label{line:bootstrap-expiry-check} \\
    \pcif t_{\mathrm{rx}} - t_{\mathrm{sign}}+\epsilon_{\mathit{u}}^{\mathrm{rx}}+\epsilon_{\mathit{g}}^\mathrm{max}>W \pcthen \pcreturn \bot \pccomment{possible signature replay} \label{line:bootstrap-replay-check} \\
    \mathit{ID} \gets \widehat{\mathit{cid}} \parallel t_{\mathrm{exp}} ;\; M \gets \widehat{\mathit{cid}} \parallel \widehat{\Pi} \parallel t_{\mathrm{sign}} \parallel t_{\mathrm{exp}} \pccomment{construct IBS identity and message} \\
    \pcif \mathsf{GG09.Verify}(\mathit{MPK},\, \mathit{ID},\, M,\, \sigma) = \bot \pcthen \pcreturn \bot \pccomment{signature invalid} \\[6pt]
    \Pi \gets \widehat{\Pi} ;\; 
    \mathit{cid} \gets \widehat{\mathit{cid}} \pccomment{bootstrap parameters} \\
    K_0 \gets \mathsf{ParseK_0}(\Pi) ;\; K_{\mathrm{anc}} \gets K_0 ;\; i_{\mathrm{anc}} \gets 0  \pccomment{initialize verification anchor}  \\
    \mathit{buf} \gets \emptyset ;\; K_0^{\mathrm{next}} \gets \bot ;\; {\mathit{sp}}^\mathrm{next} \gets 0 \\
    \pcreturn \top
  }

  \medskip
  \setcounter{pclinenumber}{0}%

  \procedure[linenumbering, skipfirstln, codesize=\small]{$\mathsf{TF5.VerifySIB1}(P,\; t_{\mathrm{rx}},\; \epsilon_{\mathit{u}}^{\mathrm{rx}},\; \epsilon_{\mathit{g}}^\mathrm{max})$}{%
    \textbf{Input:}\; P \text{ : received SIB1 packet};\; t_{\mathrm{rx}} \text{ : reception time};\; \epsilon_{\mathit{g}}^\mathrm{max} \pcskipln \\
    \epsilon_{\mathit{u}}^{\mathrm{rx}} = \epsilon_{\mathit{u}}(t_{\mathrm{rx}}) \text{ : UE clock error bound at } t_{\mathrm{rx}} \pcskipln \\
    \textbf{Output:}\; \top \text{ if packet successfully buffered, } \bot \text{ otherwise} \pcskipln \\
    \textbf{State:}\; \Pi,\; \mathit{cid},\; K_{\mathrm{anc}},\; i_{\mathrm{anc}},\; \mathit{buf},\; K_0^{\mathrm{next}},\; \mathit{sp}^\mathrm{next} \\
    (\widehat{M},\, i,\, \widehat{K}_{i-d},\, \widehat{K}_0^{\mathrm{next}},\,\widehat{\mathit{sp}}^\mathrm{next},\, \tau) \gets P \pccomment{$\widehat M$: SIB1 content; $\tau$: MAC tag} \\
    \pcif \mathsf{ParseCellID}(\widehat{M}) \neq \mathit{cid} \pcthen \pcreturn \bot \pccomment{cell ID mismatch} \\
    D_t \gets \epsilon_{\mathit{u}}^{\mathrm{rx}}+\epsilon_{\mathit{g}}^\mathrm{max} \\
    \pcif \mathsf{SafePacketTest}(t_{\mathrm{rx}},\, i,\, \Pi,\, D_t) = \bot \pcthen \pcreturn \bot \pccomment{cf.\ App.~\ref{app:safe-packet-test}} \\
    \pcif i - d > i_{\mathrm{anc}} \pcthen \\
    \pcind \pcif F^{\,(i - d) - i_{\mathrm{anc}}}(\widehat{K}_{i-d}) \neq K_{\mathrm{anc}} \pcthen \pcreturn \bot \pccomment{key verification failed} \\
    \mathit{buf} \gets \mathit{buf} \cup \{(\widehat{M},\, i,\, \widehat{K}_0^{\mathrm{next}},\, \widehat{\mathit{sp}}^{\mathrm{next}},\, \tau)\} \\
    \pcif i - d \leq i_{\mathrm{anc}} \pcthen \pcreturn \top \pccomment{disclosed key already seen; buffered only} \\
    K_{\mathrm{anc}} \gets \widehat{K}_{i-d} ;\; i_{\mathrm{anc}} \gets i - d \pccomment{advance trust anchor} \\[6pt]
    \pcwhile \exists\, \mathit{entry}_j \in \mathit{buf} \text{ with } j \leq i_{\mathrm{anc}} \pcdo \pccomment{$j$: entry's interval} \\
    \pcind (\widehat{M},\, j,\, \widehat{K}_0^{\mathrm{next}},\, \widehat{\mathit{sp}}^{\mathrm{next}},\, \tau) \gets \mathit{entry}_j \\
    \pcind \mathit{buf} \gets \mathit{buf} \setminus \{\mathit{entry}_j\} \pccomment{remove eligible entry} \\
    \pcind K_j \gets F^{\,i_{\mathrm{anc}} - j}(K_{\mathrm{anc}}) \pccomment{recover $K_j$; tolerates skips} \\
    \pcind \pcif \mathrm{MAC}(F'(K_j),\, \widehat{M} \parallel \widehat{K}_0^{\mathrm{next}} \parallel \widehat{\mathit{sp}}^{\mathrm{next}}) = \tau \pcthen \\
    \pcind \pcind \text{accept } \widehat{M} ;\; K_0^{\mathrm{next}} \gets \widehat{K}_0^{\mathrm{next}} ;\; \mathit{sp}^\mathrm{next} \gets \widehat{\mathit{sp}}^{\mathrm{next}} \pccomment{accept buffered packet} \label{line:accept-buffered-packet} \\
    \pcreturn \top
  }
\end{algorithm}

Algorithm~\ref{alg:ue-procedures} summarizes the UE-side authentication procedures using \sysabbrv{}.
We describe only \sysabbrv{}-specific elements here; see \S\ref{sec:background:tesla} for more details on TESLA operation.
Each gNB's IBS identity is $\mathit{ID} = \mathit{cid} \parallel t_{\mathrm{exp}}$, binding the Cell ID $\mathit{cid}$ with the signing key's expiration time; a short validity period (see \S\ref{sec:discussion:key-validity}) obviates UE-side revocation checking.
$\mathsf{TF5.Bootstrap}$ verifies the key validity and freshness (lines~\ref{line:bootstrap-expiry-check}--\ref{line:bootstrap-replay-check}) before accepting the message and bootstrapping the chain commitment and TESLA parameters.
$\mathsf{TF5.VerifySIB1}$ caches the last verified key as trust anchor $K_{\mathrm{anc}}$ to avoid full chain traversal, and stores the commitment to the next chain, $K_0^\mathrm{next}$, for seamless chain transitions (see \S\ref{sec:design:chain-renewal}).
``Accept $\widehat{M}$'' in line~\ref{line:accept-buffered-packet} indicates that the UE successfully verified the buffered SIB1 content and continues to process it as normal.
$\mathsf{SafePacketTest}$ is detailed in Appendix~\ref{app:safe-packet-test}.

\subsection{TESLA Timing and Parameters}
\label{sec:design:timesync}

TESLA requires each receiver to know an upper bound $D_t$ on the sender's clock advance relative to its own (\S\ref{sec:background:tesla}).
5G already has a built-in synchronization mechanism: the gNB periodically transmits the Synchronization Signal Block (SSB)~\cite{3gpp.38.213}, which the UE uses to synchronize its own clock.
However, the SSB timing is not authenticated, so a malicious gNB can forge it to mislead the UE's clock.
Therefore, \sysabbrv{} instead uses the gNB and UE's own local clocks synchronized using GPS, for both TESLA time synchronization and signature replay protection.

In \sysabbrv{}, both the gNB and UE are required to be able to calculate their own clock error bounds relative to GPS time at any time $t$ (the method is detailed in Appendix~\ref{app:clock-error-bound}).
The gNB always keeps its clock error within a fixed bound, $\epsilon_{\mathit{g}}^\mathrm{max}$, as assumed in \ref{sec:threat-model:assumptions}.
We set $\epsilon_{\mathit{g}}^\mathrm{max}=1.5\,\mu\mathrm{s}$, which is the maximum allowed error for gNBs in the NR TDD radio frame synchronization requirement~\cite{itu.G8271.2020}.
The UE does not have to tightly synchronize its clock to GPS time at all times; still, before performing any \sysabbrv{} operations, it must calculate its maximum clock error,  $\epsilon_{\mathit{u}}(t)$.
Only when UE's clock drifts too far from GPS time, it has to turn on GPS to re-synchronize to reduce its error. After re-synchronization, the UE can turn off GPS to save power.

For TESLA safe packet test~(Appendix~\ref{app:safe-packet-test}), we use $D_t=\epsilon_{\mathit{u}}(t_{\mathrm{rx}})+\epsilon_{\mathit{g}}^\mathrm{max}$, where $t_{\mathrm{rx}}$ is the time at which the UE receives the packet.
We set the TESLA interval to $T_{\mathrm{int}} = 160\,\text{ms}$, matching SIB1 broadcast periodicity~\cite{3gpp.38.331}, and the key disclosure delay $d=1$ to minimize verification latency.\footnote{RFC 4082 notes that $d = 1$ ``does not work'' for packets near interval boundaries~\cite{RFC4082-TESLA}; this does not apply to \sysabbrv{} because SIB1 broadcasts have a fixed schedule, and we can perfectly align them with TESLA intervals to avoid near-boundary packets.}

\subsection{Chain Renewal} \label{sec:design:chain-renewal}

TESLA uses a finite key chain of length $N$; once exhausted, the gNB must switch to a new chain.
Without any renewal mechanism, each chain transition would require UEs to perform \textsf{TF5.Bootstrap} again.
\sysabbrv{} supports chain renewal by embedding a \emph{next chain commitment} $K_0^{\mathrm{next}}$ in every SIB1 extension (Table~\ref{tab:wire-format}).
The gNB generates the next chain in advance and includes its root commitment, $K_0^{\mathrm{next}}$, in the current chain's SIB1 broadcasts, allowing UEs to seamlessly transition to the next chain without re-verifying a signature.
$K_0^{\mathrm{next}}$ is also authenticated by the current chain's TESLA MAC, so any UE that has verified at least one SIB1 from the current chain naturally bootstraps the next chain.

A $\mathit{sp}^\mathrm{next}$ flag is included along with $K_0^{\mathrm{next}}$ to indicate whether the next chain's parameters differ from the current chain's. $\mathit{sp}^\mathrm{next}=0$ (same parameters) means that the next chain inherits all current parameters and starts immediately after the last interval, so the UE can proceed to the next chain without re-bootstrapping. $\mathit{sp}^\mathrm{next}=1$ (different parameters) indicates that one or more parameters have changed, and the UE must re-bootstrap via a fresh \customsib{}.
If the UE has been idle for the whole current chain and has not obtained $K_0^{\mathrm{next}}$, it may have to re-bootstrap when the next chain starts.
To avoid this, \sysabbrv{} allows the UE to wake up several intervals before chain exhaustion to receive and verify at least one SIB1 and obtain $K_0^{\mathrm{next}}$.
The cost of this ``wakeup'' is evaluated in Section~\ref{sec:eval:trace-cost}.

\section{Security Analysis}
\label{sec:security}

We formally verify \sysabbrv{}'s security goals (\S\ref{sec:threat-model:goals}) using the Tamarin Prover~\cite{meierTAMARINProverSymbolic2013} under a Dolev--Yao adversary capable of compromising individual gNB signing keys.
We consulted the TESLA proof of Cremers et al.~\cite{cremersSubtermBasedProofTechniques2023} during model development.
The complete Tamarin model is available at \url{https://github.com/ssubinsong/tf5-tamarin-proof}.

Our Tamarin model employs the following abstractions.
Following prior work~\cite{meierAdvancingAutomatedSecurity2013,cremersSubtermBasedProofTechniques2023}, we represent time through symbolic trace ordering.
The receiver's signing key validity, signature freshness, and safe packet checks are enforced through restrictions on the ordering of events.
The model fixes $d=1$ and simplifies the buffer processing in $\mathsf{TF5.VerifySIB1}$ (Algorithm~\ref{alg:ue-procedures}) by allowing each UE to buffer and verify at most one SIB1 packet per interval.
Cryptographic primitives (e.g., GG09 IBS and hash functions) are idealized and treated as black boxes, while chain renewal is modeled exclusively for the case in which the parameters remain unchanged ($\mathit{sp}^\mathrm{next}{=}0$).

\begin{table}[tbp]
  \centering
  \caption{\sysabbrv{}'s security goals and auxiliary properties mapped to verified Tamarin lemmas.
    Guarantees apply to gNBs with uncompromised signing keys.}
  \label{tab:tamarin}
  \small
  \begin{tabular}{@{}p{0.28\textwidth}p{0.68\textwidth}@{}}
    \hline
    \textbf{Property} & \textbf{Tamarin Lemma(s)} \\
    \hline
    \customsib{} source authentication \& integrity
      & \texttt{bootstrap\_authentic} --- verified \customsib{} was sent by a legitimate gNB with matching content \\
    \hline
    SIB1 source authentication \& integrity
      & \texttt{sib1\_authenticity} --- verified SIB1 was sent by a legitimate gNB with matching content \\
    \hline
    \customsib{} replay protection
      & \texttt{bootstrap\_signature\_freshness} --- verified bootstrap was received before signature expiry \\
    \hline
    SIB1 replay protection
      & \texttt{sib1\_freshness} ---
        verified SIB1 was received before its TESLA interval expired \\
    \hline
    TESLA key secrecy
      & \texttt{knows\_only\_expired\_chain\_keys} --- non-root chain keys remain secret until their intervals expire \\
    \hline
    gNB signing key secrecy
      & \texttt{gnb\_signing\_key\_secrecy} --- a gNB signing key remains secret \\
    \hline
    gNB key validity conformance
      & \texttt{bootstrap\_verification\_requires\_valid\_signing\_key} --- verified bootstrap was received while the gNB signing key was valid \\
    \hline
    Chain renewal authenticity
      & \texttt{renewal\_commitment\_authenticity},
        \texttt{renewal\_requires\_authenticated\_commitment} --- UE renewal uses only the successor commitment ($K_0^\mathrm{next}$) designated by the gNB for the current chain \\
    \hline
  \end{tabular}
\end{table}

Table~\ref{tab:tamarin} maps the security goals (\S\ref{sec:threat-model:goals}) and auxiliary properties to their corresponding Tamarin lemmas. We now discuss how the lemmas guarantee each security goal, assuming that the gNB's signing key remains uncompromised.

\paragraph{Source authentication.}
\texttt{sib1\_authenticity} (Appendix~\ref{app:sib1-authenticity}) guarantees that every accepted SIB1 corresponds to an earlier transmission by the legitimate gNB.
An SIB1 fabricated by a rogue gNB would lack such a transmission, so accepting it would contradict the lemma.

\paragraph{Message integrity.}
\texttt{sib1\_authenticity} requires the transmitted and accepted SIB1 content to match.
If an attacker changes SIB1 content from $m$ to $m'$, accepting $m'$ requires a prior transmission of $m'$ by the legitimate gNB.
Without such a transmission, accepting the modified content would contradict the lemma.

\paragraph{Replay protection.}
\texttt{sib1\_freshness} (Appendix~\ref{app:sib1-freshness}) shows that accepted SIB1 packets were received before their original TESLA intervals expired.
A replay received after its original interval expires would violate this ordering if accepted, even with a valid MAC.

\paragraph{Sanity checks and auxiliary properties.}
Additionally, we verify eleven \texttt{exists-trace} sanity lemmas (not in the table) covering normal protocol runs and adversarial behavior.
These checks ensure that the security lemmas do not hold vacuously due to the absence of relevant executions.
Auxiliary security properties are also verified, including TESLA/signing key secrecy, gNB signing key validity conformance, and chain renewal authenticity (Table~\ref{tab:tamarin}).
The complete model contains further details.

\section{Evaluation}
\label{sec:evaluation}

\subsection{Computational Overhead} \label{sec:eval:compute}

\subsubsection{Evaluation setup.}
We compare \sysabbrv{} against seven baselines: certificate-based schemes ($\mathsf{Cert}_\text{ECDSA}$, $\mathsf{Cert}_\text{EdDSA}$), pairing-free IBS (GG09~\cite{GG09}, \schnorrhibs{}~\cite{singlaLookYouLeap2021}), and pairing-based schemes (Hes03~\cite{HessIBS}, BLMQ~\cite{BLMQ}, and $\mathsf{Cert}_\text{BLS}$\footnote{We develop a scheme that uses BLS~\cite{BLS} for both certificate and message signing and name it $\mathsf{Cert}_\text{BLS}$. BLS is not a standard algorithm for PKI certificates---still, we consider it because it has a compact signature size and is thus suitable for size-constrained SI messages.}).
These baselines align with 3GPP TR~33.809~\cite{3gpp.33.809}, which proposes certificate-based schemes, IBS (e.g., ECCSI~\cite{rfc6507}, SM9~\cite{ISO14888-3-2018}), and BLS~\cite{BLS} as SI authentication candidate methods.
We evaluate the computational overhead of each scheme across three roles: \emph{core} (gNB key extraction/certificate issuance), \emph{gNB} (signing), and \emph{UE} (signature/MAC verification).
\schnorrhibs{}~\cite{singlaLookYouLeap2021} employs a 3-level hierarchy (PKG, AMF, gNB) for GG09 IBS in 5G; we adopt its 2-level variant (PKG, gNB) as the ``GG09'' baseline in the experiments.

All Schnorr-based schemes (\sysabbrv{}, $\mathsf{Cert}_\text{EdDSA}$, GG09, \schnorrhibs{}) are implemented on the Edwards25519 curve\footnote{\schnorrhibs{} originally uses FourQ~\cite{FourQ}; we reimplement it on Edwards25519 for fair comparison.} using \libsodium{}~\cite{LibsodiumLibrary}.
Pairing-based schemes (Hes03, BLMQ, $\mathsf{Cert}_\text{BLS}$) use BLS12-381 via \blst{}~\cite{BlstLibrary}, while $\mathsf{Cert}_\text{ECDSA}$ uses NIST P-256 through OpenSSL~3.0+~\cite{OpensslLibrary}.
OpenSSL is also used for SHA-256 and HMAC.
We apply standard optimizations to all schemes, including offline/online signing decomposition and precomputed \textit{MPK} tables.
Each operation is measured over 1,000 iterations after 10 warmup runs, and we report the average computation time in microseconds.
We will make our evaluation code publicly available in the camera-ready version.

We benchmark all three roles on a \emph{desktop} (Intel Core i5-14600KF, 5.3\,GHz, 32\,GB RAM) and use a \emph{Raspberry Pi (RPi)~4B} (Cortex-A72 @ 1.8\,GHz, 8\,GB RAM) for UE-side evaluation to simulate power-constrained UEs.
To estimate the performance across diverse UE classes, we scale the RPi measurements to three platforms using benchmarks from \ebacs{}~\cite{eBACS} and \coremark{}~\cite{eembcCoreMark}: smartphone (Cortex-A76 @ 2.42\,GHz), V2X (Cortex-A53 @ 1.0\,GHz), and IoT (Cortex-M33 @ 64\,MHz).
The IoT platform has a different ISA (32-bit ARMv8-M) from RPi (64-bit ARMv8-A) and the projections are likely to be underestimates; further details are in Appendix~\ref{appendix:scaling}.

\begin{table}[tbp]
    \centering
    \caption{Computational evaluation result in $\mu$s. Core and gNB operations are measured on a desktop; UE-side verification are measured on both a desktop and a RPi~4B. Results measured on a RPi~4B are subsequently projected to smartphone, V2X, and IoT platforms.
    Sign-off/on = gNB signing offline (precomputation)/online.
    Bold indicates the three lowest values in each column.}
    \label{tab:eval-compute-results}
    \begin{threeparttable}
        {
        \begin{tabular}{lrrrrrrrr}
            \hline
            & \multicolumn{1}{c}{\textbf{Core}} & \multicolumn{2}{c}{\textbf{gNB}} & \multicolumn{5}{c}{\textbf{UE Verification}} \\
            \cmidrule(lr){2-2} \cmidrule(lr){3-4} \cmidrule(lr){5-9}
                          & \textbf{Extract\tnote{$\dagger$}} & \textbf{Sign-off} & \textbf{Sign-on}
                          & \raisebox{0.1ex}{\textbf{\scriptsize Desktop}} & \textbf{RPi 4B} & \textbf{Phone} & \textbf{V2X} & \textbf{IoT} \\
            \hline
            \sysabbrv{} bootstrap & \textbf{10.81} & \textbf{10.35} & \textbf{0.33} & \textbf{41.95} & \textbf{453.93} & \textbf{278} & \textbf{1{,}212} & \textbf{17{,}873}\tnote{$\triangle$} \\
            \sysabbrv{} MAC\tnote{$\ddagger$} & --- & \textbf{0.17} & 0.67 & \textbf{0.80} & \textbf{4.02} & \textbf{0.53} & \textbf{10.3} & \textbf{158}\tnote{$\triangle$} \\
            \hline
            $\mathsf{Cert}_\text{ECDSA}$ & 14.27 & 16.45 & \textbf{0.49} & 87.77 & 611.48 & 340 & 1{,}637 & 24{,}077\tnote{$\triangle$} \\
            $\mathsf{Cert}_\text{EdDSA}$ & \textbf{11.06} & \textbf{10.36} & \textbf{0.50} & 59.69 & 641.68 & 393 & 1{,}714 & 25{,}266\tnote{$\triangle$} \\
            \hline
            $\mathsf{Cert}_\text{BLS}$ & 103.11 & --- & 103.11 & 805.81 & 9{,}374.10 & 5{,}265\tnote{$\ast$} & 29{,}877\tnote{$\ast$} & 369{,}097\tnote{$\triangle$} \\
            Hes03 & 93.44 & 368.57 & 57.93 & 526.12 & 6{,}117.40 & 3{,}436\tnote{$\ast$} & 19{,}498\tnote{$\ast$} & 240{,}867\tnote{$\triangle$} \\
            BLMQ  & 58.10 & 368.71 & 57.42 & 662.32 & 7{,}708.22 & 4{,}330\tnote{$\ast$} & 24{,}568\tnote{$\ast$} & 303{,}505\tnote{$\triangle$} \\
            \hline
            GG09 & \textbf{10.73} & 10.42 & \textbf{0.50} & \textbf{41.91} & \textbf{451.56} & \textbf{276} & \textbf{1{,}206} & \textbf{17{,}780}\tnote{$\triangle$} \\
            \schnorrhibs{} & 21.64 & \textbf{10.36} & \textbf{0.50} & 74.26 & 802.30 & 491 & 2{,}143 & 31{,}590\tnote{$\triangle$} \\
            \hline
        \end{tabular}
        }
        \begin{tablenotes}\footnotesize
            \item[$\dagger$] Key extraction for IBS schemes; certificate issuance for \textsf{Cert} schemes.
            \item[] Smartphone, V2X, IoT values \emph{without} markers use \ebacs{}~\cite{eBACS} scaling (high confidence).
            \item[$\ast$] Uses \coremark{}/MHz~\cite{eembcCoreMark} scaling (medium confidence---same ISA).
            \item[$\triangle$] Uses \coremark{}/MHz~\cite{eembcCoreMark} scaling (low confidence---cross-ISA).
            \item Standard deviations for desktop and RPi~4B measurements: For values ${\geq}1$\,$\mu$s, standard deviations are ${<}6\%$ of the mean, except for RPi MAC verification which is ${<}12\%$ of the mean. For values ${<}1$\,$\mu$s, standard deviations are below ${<}14\%$ of the mean.
        \end{tablenotes}
    \end{threeparttable}
\end{table}

\subsubsection{Results.}
Table~\ref{tab:eval-compute-results} summarizes the benchmark results.
Since \sysabbrv{} and GG09 use the same signing algorithm and differ only in their payloads, their performance is nearly identical across all roles and platforms.
Overall, pairing-based schemes ($\mathsf{Cert}_\text{BLS}$, Hes03, BLMQ) are 1--2 orders of magnitude slower than non-pairing schemes.
All non-pairing schemes take less than 1\,$\mu$s for online signing at the gNB, and \sysabbrv{}'s MAC calculation is comparable at 0.67\,$\mu$s.
Excluding the \sysabbrv{} MAC, GG09 and \sysabbrv{}'s bootstrap verification have the lowest UE verification cost, on both desktop and RPi.
While $\mathsf{Cert}_\text{EdDSA}$ verification outperforms $\mathsf{Cert}_\text{ECDSA}$ on the desktop, the opposite result occurs on the RPi because OpenSSL's optimized assembly for P-256 is unavailable for ARMv8-A.
The MAC verification cost is 0.80\,$\mu$s on the desktop ($52\times$ faster than GG09), 4.02\,$\mu$s on the RPi ($112\times$), and 0.53\,$\mu$s on the smartphone ($521\times$).

We also evaluate the hash chain traversal cost of TESLA.
To quantify this cost, we measure the computation time for $x$ consequent SHA-256 operations on both desktop and RPi, where $x \in \{1, \dots, 50{,}000\}$.
Fitting the results to a linear regression model, we find that the desktop and RPi show near-perfect linearity ($R^2 \approx 1$) with slopes of $0.14$\,\textmu{}s and $0.61$\,\textmu{}s per $\Delta x$, respectively.
Predicted slopes for smartphone, V2X, and IoT are $0.08$, $1.58$, and $24.2$\,\textmu{}s/$\Delta x$.
Hashing on smartphones is fast due to ARMv8 Crypto Extensions, which are prevalent in modern mobile SoCs~\cite{kolblPuttingWingsSPHINCS2018}.

\subsection{Communication and Storage Overhead}
\label{sec:eval:comm-storage}

\begin{table}[tbp]
    \centering
    \caption{Air-interface communication overhead (bytes). per-cell = data that is cacheable while the UE stays in the same cell; per-SIB1 = data that changes on every SIB1 broadcast. \textsf{Cert} schemes report the sum of raw field sizes, not the X.509 DER encoding.}
    \label{tab:eval-comm-overhead}
    \begin{tabular}{lrrr}
        \hline
        \textbf{Scheme}                    & \textbf{per-cell} & \textbf{per-SIB1} & \textbf{Total} \\
        \hline
        \sysabbrv{}      & 134              & 53               & 187             \\
        \hline
        $\mathsf{Cert}_\text{ECDSA}$                         & 100              & 67               & 167            \\
        $\mathsf{Cert}_\text{EdDSA}$                         & 99               & 67               & 166            \\
        \hline
        $\mathsf{Cert}_\text{BLS}$                            & 147              & 51               & 198            \\
        Hes03, BLMQ                         & 3                & 83               & 86             \\
        \hline
        GG09                              & 35               & 67               & 102            \\
        \schnorrhibs{}                     & 77               & 67               & 144            \\
        \hline
    \end{tabular}
\end{table}

Table~\ref{tab:eval-comm-overhead} summarizes the air-interface overhead for each scheme.
We normalize timestamp and key validity fields in all schemes to 3\,B. For IBS schemes, we do not count the cell ID field into message size, as it is already present in original SIB1 content and can be derived from there.

Since per-cell data (e.g., certificates, IBS parameters) can be cached at the UE while it stays in the same cell, the critical metric is per-SIB1 overhead (i.e., the data included in every SIB1 broadcast).
\sysabbrv{}'s per-SIB1 overhead (53\,B) is lower than all baselines except $\mathsf{Cert}_\text{BLS}$ (51\,B). However, $\mathsf{Cert}_\text{BLS}$ has a high verification cost~(\S\ref{sec:eval:compute}), which is undesirable for UEs with limited computational resources.
Hes03 and BLMQ achieve the lowest total overhead ($3+83=86$\,B) but the highest per-SIB1 cost (83\,B).
The maximum capacity of SIB1 is 372\,B~\cite{3gpp.38.331}, and a typical SIB1 message occupies ${\sim}100$\,B,\footnote{Measured from a default config in OpenAirInterface (OAI)~\cite{OpenAirInterface}.} there is only ${\sim}272$\,B left for authentication data.
\sysabbrv{}'s 53\,B per-SIB overhead comfortably fits within this budget.
All certificate schemes also fit within the SIB1 budget since we measured the raw field sizes.
However, DER-encoded X.509 certificates increase per-cell overhead to 317\,B for ECDSA and 252\,B for EdDSA,\footnote{Measured using OpenSSL~3.0+~\cite{OpensslLibrary}} exceeding the SIB1 capacity and requiring a separate SIB.

\begin{table}[tbp]
    \caption{Storage overhead per scheme.}
    \label{tab:eval-storage}
    \begin{subtable}[t]{0.5\textwidth}
        \centering
        \caption{Key sizes (bytes). For $\mathsf{Cert}$ schemes, $\mathit{MPK}$ and $\mathit{msk}$ are the public and private keys of the CA, respectively.}
        \label{tab:eval-key-storage}
        \begin{tabular}{lrrrr}
            \hline
            \textbf{Scheme} & \scriptsize \textbf{$\mathit{MPK}$} & \scriptsize \textbf{$msk$} & \scriptsize \textbf{$PK_\mathsf{gnb}$} & \scriptsize \textbf{$sk_\mathsf{gnb}$} \\
            \hline
            \scriptsize \sysabbrv{},  \schnorrhibs{},          & \multirow{2}{*}{32} & \multirow{2}{*}{32} & \multirow{2}{*}{32} & \multirow{2}{*}{32} \\
            \scriptsize \quad GG09, $\mathsf{Cert}_\text{EdDSA}$      &    &    &    &    \\
            \hline
            $\mathsf{Cert}_\mathrm{ECDSA}$              & 33          & 32          & 33              & 32              \\
            $\mathsf{Cert}_\mathrm{BLS}$                & 96          & 32          & 96              & 32              \\
            Hes03, BLMQ            & 96          & 32          & ---             & 48              \\
            \hline
        \end{tabular}
    \end{subtable}
    \hfill
    \begin{subtable}[t]{0.49\textwidth}
        \centering
        \caption{\sysabbrv{} runtime state sizes (bytes).}
        \label{tab:eval-storage-tf5-extra}
        \begin{tabular}{llr}
            \hline
        \raisebox{0.1ex}{\scriptsize \textbf{Entity}} & \textbf{Item} & \textbf{Size} \\
            \hline
            gNB & TESLA params ($\Pi$)          & 11            \\
                & TESLA key chain           & $16 \cdot N$ \\
            \hline
            UE  & TESLA params ($\Pi$)          & 11            \\
                & $\mathit{cid}$ (36\,b), $\mathit{sp}^\mathrm{next}$ (1\,b) & 5 \\
                & Trust anchor key ($K_{\mathrm{anc}}$) & 16         \\
                & Trust anchor idx. ($i_{\mathrm{anc}}$) & 4         \\
                & Pending SIB1 buffer ($\mathit{buf}$) & 392   \\
                & Next-chain comm.\ ($K_0^{\mathrm{next}}$) & 16         \\
            \hline
        \end{tabular}
    \end{subtable}
\end{table}

Table~\ref{tab:eval-key-storage} summarizes the key sizes for each scheme.
$\mathit{MPK}$s of all schemes are small (32--96\,B), so they can be easily pre-provisioned to UEs' SIM cards.
Table~\ref{tab:eval-storage-tf5-extra} details the runtime state sizes for \sysabbrv{}.
The gNB's TESLA key chain ($16 \times N$\,B) is the dominant factor, requiring ${\approx}32$\,KB when $N = 2{,}000$, while the UE state totals 444\,B.

\subsection{Implementation and End-to-End Latency}
\label{sec:eval:impl}
\label{sec:eval:latency}

We prototype \sysabbrv{} on the OpenAirInterface (OAI)~\cite{OpenAirInterface} 5G stack. We implement GG09 IBS using Ed25519 via \libsodium{}~\cite{LibsodiumLibrary}, while SHA-256 hash and HMAC operations use OpenSSL 3.0+~\cite{OpensslLibrary}.
We measure delay from SIB1 decoding to camp completion across 50 runs on a desktop. 
The system is deployed via Docker using OAI's \texttt{vrtsim}~\cite{oai-vrtsim} at a 1.0 timescale. We evaluate two scenarios: a new UE ($\mathsf{TF5.Bootstrap}$ + $\mathsf{VerifySIB1}$) and a returning UE ($\mathsf{TF5.VerifySIB1}$). The \customsib{} broadcast period is set to 160\,ms.

\begin{table}[tbp]
\centering
\caption{End-to-end latency (SIB1 decoded $\to$ camp) on the OAI prototype in milliseconds (50 runs).}
\label{tab:eval-latency}
\begin{tabular}{l r r r}
\hline
\textbf{Scenario} & \textbf{Mean} & \textbf{Stddev} & \textbf{Median} \\
\hline
New UE            & 289.0 & 54.7 & 290.1 \\
\quad SIB1 decoded $\to$ IBS verified          & 117.9 & 54.2 & --- \\
\quad IBS verified $\to$ camp     & 171.1 & \phantom{0}4.4 & --- \\
\hline
Returning UE (SIB1 decoded $\to$ camp)                     & 108.8 & 59.3 & 159.6 \\
\hline
\end{tabular}
\end{table}

Table~\ref{tab:eval-latency} reports the results.
For a new UE, it takes 117.9\,ms from SIB1 decoding to IBS verification and 171.1\,ms for IBS verification to camp completion, totaling 289.0\,ms.
For a returning UE, the latency is dominated by the wait for TESLA key disclosure; the total delay is 108.8\,ms, which is below $T_{\mathrm{int}}=160\,ms$.
Standard deviations are high in both scenarios because the UE enters the cell at random times relative to the \customsib{} broadcast and TESLA key disclosure intervals.
As we discussed in \S\ref{sec:design:motivation}, UEs are likely to have enough idle time after acquiring SIB1 in most cases; therefore, the TESLA delay has minimal impact on user experience.

\subsection{Trace-Driven Cost Analysis}
\label{sec:eval:trace-cost}

In this section, we refer to a UE's transition from \rrcconnected{} to \rrcidle{} as an \emph{idle return}.

We collect two real-world traces and run simulations to evaluate the daily UE-side verification cost of \sysabbrv{} relative to a signature-only baseline (GG09).
Each trace was collected using a custom Android application that monitors modem-level RRC state transitions via the radio logcat buffer.
We also log service state updates and cell ID changes (via \texttt{TelephonyCallback}).
Based on our analysis in \S\ref{sec:design:motivation}, we focus on two common events that trigger SIB1 reacquisition: cell reselection and idle return. 
Handovers do not trigger broadcast SIB1 acquisition, but we log them to track cell transitions.

We collected traces from two devices: \traceD{} (Durham, NC, USA; car-dominated; ${\approx}371$\,h; 590 reselections, 1{,}079 handovers, 1{,}763 idle returns) and \traceS{} (Seoul, South Korea; dense urban; ${\approx}376.5$\,h; 2{,}008 reselections, 2{,}825 handovers, 6{,}481 idle returns).
Overall, \traceS{} exhibits $3.3{\times}$ more events than \traceD{}.

Based on these traces, we develop a discrete event simulator to estimate verification costs.
Under a signature-only baseline, each SIB1 reception requires one signature verification ($C_{\mathrm{sig}}$).
For \sysabbrv{}, costs vary by event type:
\begin{itemize}
\item SIB1 verification on \emph{cell reselection}: $C_{\mathrm{sig}} + i \cdot C_{\text{hash}} + C_{\text{mac}}$,
where $i$ is the current interval index.
\item SIB1 verification on \emph{idle return}: $(i - i_\mathrm{anc}) \cdot C_{\text{hash}} + C_{\text{mac}}$, where $i_\mathrm{anc}$ is the last verified key's interval index.
\item Chain transition: 0 if $K_0^{\mathrm{next}}$ is pre-acquired via chain renewal; otherwise, $C_{\mathrm{sig}}$.
\end{itemize}

For all cost calculations in this section, we apply the smartphone values from \S\ref{sec:eval:compute} to $C_{\mathrm{sig}}$, $C_{\text{hash}}$, and $C_{\text{mac}}$.

\subsubsection{Chain length sensitivity}

\begin{table}[tbp]
  \caption{Trace-driven cost analysis results. $\downarrow$ = cost reduction (\%) of \sysabbrv{} relative to GG09; negative values indicate \sysabbrv{} is costlier. Bold marks the best reduction per column. Note: GG09 baseline costs are \traceD{} $= 42.1$\,ms/day, \traceS{} $= 149.6$\,ms/day.}
    \label{tab:eval-trace-combined}
    \centering
    \begin{subtable}[t]{0.48\textwidth}
        \centering
        \caption{Chain length ($N$) sensitivity. C.D.\ = chain duration in seconds ($N \times T_{\mathrm{int}}$).}
        \label{tab:eval-trace-n-sensitivity}
        \begin{tabular}{rr rr rr}
            \hline
            & & \multicolumn{2}{c}{\traceD{}} & \multicolumn{2}{c}{\traceS{}} \\
            \cline{3-4} \cline{5-6}
            $N$ & C.D. & ms/day & $\downarrow\;\;$ & ms/day & $\downarrow\;\;$ \\
            \hline
            500             & 80        & 30.8 & 26.9      & 116.2 & 22.3 \\
            1{,}000         & 160       & 28.1 & 33.1      & 106.9 & 28.6 \\
            2{,}000         & 320       & 27.7 & \textbf{34.0}      & 105.2 & \textbf{29.7} \\
            3{,}000         & 480       & 29.0 & 31.1      & 108.7 & 27.4 \\
            4{,}000         & 640       & 30.6 & 27.2      & 110.2 & 26.4 \\
            10{,}000        & 1{,}600   & 42.1 & $-$0.2    & 128.7 & 14.0 \\
            \hline
        \end{tabular}
    \end{subtable}%
    \hfill
    \begin{subtable}[t]{0.51\textwidth}
        \centering
        \caption{\sysabbrv{} feature evaluation at $N = 2{,}000$.}
        \label{tab:eval-trace-feature-eval}
        \begin{tabular}{l rr rr}
            \hline
            & \multicolumn{2}{c}{\traceD{}} & \multicolumn{2}{c}{\traceS{}} \\
            \cline{2-3} \cline{4-5}
            Config & ms/day & $\downarrow\;\;$ & ms/day & $\downarrow\;\;$ \\
            \hline
            no features        & 27.7 & 34.0      & 105.2 & 29.7 \\
            SC              & 24.2 & 42.4      & 88.4  & 40.9 \\
            CR              & 23.1 & 45.0      & 77.9  & 48.0 \\
            CR+W          & 50.9 & $-$20.9   & 71.6  & 52.2 \\
            CR+SC         & 18.9 & \textbf{55.0}      & 59.0  & 60.6 \\
            CR+W+SC     & 45.8 & $-$8.9    & 52.6  & \textbf{64.9} \\
            \hline
        \end{tabular}
    \end{subtable}
\end{table}

Table~\ref{tab:eval-trace-n-sensitivity} reports the average daily verification cost across varying $N$ for \sysabbrv{}, without chain renewal or state caching.
While increasing $N$ reduces re-bootstrapping frequency, it increases the hash chain traversal cost.
$N = 2{,}000$ yields the highest cost reduction in both \traceD{} (34.0\%) and \traceS{} (29.7\%), compared to the GG09 baseline.
Beyond $N = 2{,}000$, chain traversal cost dominates, and \sysabbrv{} gets costlier.
However, \traceS{} is more robust to large~$N$ than \traceD{} because its higher event density in an urban environment amortizes the traversal cost across more events.

\subsubsection{Feature evaluation}

Table~\ref{tab:eval-trace-feature-eval} evaluates six configurations at $N = 2{,}000$ by progressively enabling TESLA state caching (SC), chain renewal (CR), and renewal wakeup (W).
State caching allows a UE re-entering a previously visited cell (e.g., $a \to b \to a$) to resume from its cached anchor $K_{\mathrm{anc}}$, effectively mitigating ping-pong reselections at a cell edge. Details for chain renewal and wakeup are in \S~\ref{sec:design:chain-renewal}.
Both SC and CR consistently and independently improve costs: SC provides an 8--13\,pp gain compared to GG09 across all feature mixes (no features$\to$SC, CR$\to$CR+SC, CR+W$\to$CR+W+SC), while CR adds 11--20\,pp (no features$\to$CR, SC$\to$CR+SC).

However, the wakeup mechanism (W) is highly environment-sensitive.
In \traceD{}, W worsens the reduction rate by ${\approx}65$\,pp, while in \traceS{}, it improves the reduction rate by ${\approx}4$\,pp.
We attribute this to the difference in event density between the two traces.
Without W, a UE that does not acquire SIB1 for multiple chain lifetimes pays only $C_{\mathrm{sig}}$ when it acquires SIB1 for the next time.
However, with W, the UE must wake for each intermediate chain, so the accumulated hash traversal cost can easily exceed $C_{\mathrm{sig}}$.
Since \traceD{} has a lower event density, it is more likely to have long gaps between SIB1 acquisitions---therefore, W is less effective in \traceD{}.
The optimal configuration for \traceD{} is CR+SC ($18.9$\,ms/day, $55.0\%$ lower than GG09), while \traceS{} performs best with CR+W+SC ($52.6$\,ms/day, $64.9\%$ lower).
The wakeup mechanism does not need gNB coordination; UEs can adaptively toggle it based on their traffic patterns and environment to achieve best performance.

\section{Discussion} \label{sec:discussion}

\subsubsection{Roaming Scenarios.}
\label{sec:discussion:roaming}

A roaming UE needs a valid \mpk{} of the visited operator to use \sysabbrv{}.
To achieve this, a UE can obtain the visited operator's \mpk{} out-of-band (e.g., over Wi-Fi).
The home operator's PKG, which the UE already trusts, must sign this \mpk{}.
After verifying the signature, the UE can proceed with \sysabbrv{} bootstrap.

If no out-of-band channel is available at the time of initial access, the UE cannot use \sysabbrv{} to verify the visited operator's SI broadcast.
In this case, the UE has to trust the gNB on first contact without authentication.
If the gNB is legitimate and the UE successfully establishes a secure connection with the core network, the UE can then obtain the visited operator's \mpk{} and use it to verify future broadcasts.
To avoid such fallback scenarios, a user planning to travel abroad can preferably pre-download the \mpk{}s of operators in the destination country.

\subsubsection{Relay Attacks.}
\label{sec:discussion:relay}

As mentioned in Section~\ref{sec:threat-model:adversary}, \sysabbrv{} does not defend against \emph{relay attacks}: attacks where an attacker-controlled UE forwards legitimate broadcasts from a genuine gNB to a victim UE via a rogue gNB~\cite{dongEvaluatingTimeBoundedDefense2025}.
The most used countermeasure in prior work is a time-bounded defense: the gNB includes a signing timestamp and the UE rejects messages older than a window~$\Delta t$~\cite{hussainInsecureConnectionBootstrapping2019,singlaLookYouLeap2021,dongEvaluatingTimeBoundedDefense2025}.
Dong et al.~\cite{dongEvaluatingTimeBoundedDefense2025} evaluated this on an OAI testbed, separating legitimate (${\sim}3$\,ms) from relayed (${\sim}7$\,ms) messages at $\Delta t = 5$\,ms. 

\subsubsection{Signing Key Validity Period.}
\label{sec:discussion:key-validity}

Each gNB's signing key includes a short validity period in its identity: $\mathit{ID} = \mathit{cid} \| t_{\mathrm{exp}}$.
A short validity period makes explicit revocation unnecessary.
The use of such short-lived keys appears in identity-based cryptography~\cite{bonehIdentityBasedEncryptionWeil2001} and has also been adopted in \cite{singlaLookYouLeap2021,hussainInsecureConnectionBootstrapping2019}.
Key renewal cost is small: $\mathsf{GG09.Extract}$ takes ${\approx}\,11\,\mu\mathrm{s}$ per key on a desktop (Table~\ref{tab:eval-compute-results}). A PKG serving about $50{,}000$ gNBs\footnote{The number of average gNBs per operator in the United States in 2024 Q4 was approximately $46{,}700$ ($140{,}000$ gNBs nationwide in 2024 Q4~\cite{5GObservatoryReport}, and 3 MNOs~\cite{FCCCommunicationsMarketplace2024}).} finishes all extractions in $\approx 0.55$\,s on a single core, and each renewal transmits only a 64-byte key per gNB (${\approx}\,3.2$\,MB for all gNBs total).
Therefore, setting the validity period as short as ten minutes would not impose significant overhead on the PKG.

\section{Related Work}
\label{sec:related}

In this section, we discuss prior work on SI authentication and other approaches to mitigate fake base station attacks, as well as device-side detections and defenses. We also highlight concurrent work on TESLA-based SIB authentication.

\subsubsection{SI Authentication and Other Approaches.}
Hussain et al.~\cite{hussainInsecureConnectionBootstrapping2019} present an early approach to SI authentication to prevent fake base station attacks, using certificates and digital signatures. They also evaluate TESLA but consider it infeasible for SIB authentication, since it adds latency due to delayed key disclosure. We revisit this concern in \S\ref{sec:design:motivation}, and summarize their TESLA evaluation and compare it with ours in Appendix~\ref{app:prior-tesla-eval}.

3GPP TR 33.809~\cite{3gpp.33.809} discusses certificate-based signature schemes (ECDSA, RSA), identity-based and certificateless solutions (ECCSI~\cite{rfc6507}, SM9~\cite{ISO14888-3-2018}), and BLS signatures~\cite{BLS} for fake base station mitigation. It further addresses delegated signing via a Digital Signing Network Function (DSnF), whose practicality is evaluated by Gao et al.~\cite{gaoEvaluatingDelegatedDigital2021}. 3GPP TR 33.809 also mentions TESLA, noting that key commitments require authentication through an additional mechanism, such as digital signatures.

Ross et al.~\cite{rossFixingInsecureCellular2024} propose a backward-compatible, certificate-based scheme for authenticating all SIB types.
Purification et al.~\cite{wuthierBaseStationCertificate2025} pre-distribute base station certificates via a blockchain and combine location checks with signature verification for base station authentication.
To eliminate the certificate transmission overhead, Singla et al.~\cite{singlaLookYouLeap2021} introduce \schnorrhibs{}, 
a hierarchical identity-based signature (IBS) scheme built on GG09~\cite{GG09}, which Dong et al.~\cite{E2IBSPaper} subsequently optimize into \etibs{} ($2\times$ faster verification than \schnorrhibs{}).
Complementing these efforts, Dong et al.~\cite{dongEvaluatingTimeBoundedDefense2025} evaluate the effectiveness of a time-bounded defense against relay attacks on signature-based solutions.

Lotto et al.~\cite{lottoBARONBaseStationAuthentication2023} propose BARON, in which a UE registered with the core network authenticates a target base station by verifying a token from a trusted node (e.g., the core network). However, BARON does not authenticate broadcast SI.

\subsubsection{Device-Side Detections and Defenses.}
Device-side defenses detect attacks or unsafe behaviors in control plane traffic without requiring protocol changes.
PHOENIX~\cite{echeverriaPHOENIXDeviceCentricCellular2021} checks LTE control traffic against behavioral signatures, defined over relative message ordering and security state, using lightweight runtime verification.
Purification et al.~\cite{purificationFakeBaseStation2024} detect and blacklist FBSes that stall registration procedures, based on elapsed time and registration request counts.
FBSDetector~\cite{mubasshirGottaDetectemAll2025} classifies each message as benign or malicious given its surrounding context, and further learns graph patterns over message sequences to recognize multi-step attacks. 

\subsubsection{Concurrent Work.}
After completing this work, we became aware of an independently conducted concurrent work on TESLA-based SIB authentication.
Darzi et al.~\cite{darziLightweightPostQuantumAuthentication2026} propose EMULSION, a protocol for authenticating SIBs in 5G and 6G mobile networks using TESLA with post-quantum signatures.
The key differences between their work and ours are as follows.
First, Darzi et al. focus on post-quantum security and employ the MAYO-2 signature scheme~\cite{beullensMAYOPracticalPostquantum2022}, whereas \sysabbrv{} uses the GG09 IBS scheme~\cite{GG09}.
Second, in their design, the AMF in the core network is responsible for signature and TESLA key chain generation, while in \sysabbrv{}, the gNB performs both operations.
In addition, to evaluate the practical cost of using TESLA for SIB1 verification, we collect radio event traces on Android phones and, based on these traces, compare the projected cost of TESLA against a signature-only baseline (see \S\ref{sec:eval:trace-cost}).

\section{Conclusion}
\label{sec:conclusion}

We presented \sysname{} (\sysabbrv{}), a broadcast authentication protocol for 5G SIB1 that combines GG09 IBS~\cite{GG09} with TESLA~\cite{RFC4082-TESLA}.
\sysabbrv{} replaces costly per-SIB1 signature verification of prior schemes with efficient MAC checks.
We verified its security with Tamarin Prover; compared its computation, communication, and storage overhead against seven baseline schemes.
We also implemented a prototype of \sysabbrv{} on OpenAirInterface, achieving a verification latency of ${\approx}109$\,ms for returning UEs.
A trace-driven cost analysis on two SIB1 reception traces collected in real environments shows a 55--65\% reduction in daily verification cost compared to a signature-only baseline.
By demonstrating its cost-effectiveness on real-world event traces, \sysabbrv{} is a viable candidate for addressing the unresolved problem of secure 5G broadcast authentication.

\clearpage
\appendix
\renewcommand{\theHsection}{\Alph{section}}%
\section*{Supplementary Material}
\suppressfloats[t]
\section{Frequency of SIB1 Acquisition Events}
\label{app:delay-sensitive-event-frequency}

In order to justify our claim in \S\ref{sec:design:motivation} that certain events that cause SIB1 acquisition happen \emph{infrequently}, we summarize prior work that quantify the frequency of these events. The events we consider are initial registration, returning from out of service, and radio link failure (RLF). These events are delay-sensitive in \sysabbrv{} because the UE may need to connect to the gNB as soon as possible (in contrast to idle returns and cell reselections), and TESLA's delay may affect user experience.

\paragraph{Initial registration.}
Meng et al.~\cite{mengModelingGeneratingControlPlane2023} analyzed LTE signaling events from $23{,}388$ phones over a week in the US. In their collected data, the LTE Attach event only accounts for $0.1\%$ of all signaling events, whereas the S1 connection Release event (which triggers an idle return for the UE) accounts for $47.5\%$. This suggests that initial registration happens rarely compared to idle returns.
\paragraph{Returning from out of service.}
Li et al.~\cite{liNationwideStudyCellular2021} analyzed around 70 million Android phones in China for eight months in 2020 (with various generations from 2G to 5G).
They found that 95\% of the phones did not experience any out of service events during the entire eight months.
This suggests that returning from out of service is a rare event for most users.

\paragraph{Radio link failure (RLF).}
We could not find a study that directly quantifies the frequency of RLFs.
Instead, we found a work that quantifies the frequency of handover failures (which are often caused by RLFs).
Kalntis et al.~\cite{kalntisTelcoLensCountrywide2024} analyzed four weeks of handover events from a top-tier European operator among 40 million UEs (LTE and 5G NSA), and they report that the daily median handover failure rate per cell sector is $0.04\%$.

\section{TESLA Safe Packet Test and Clock Error Bounds}

\subsection{Safe Packet Test.}
\label{app:safe-packet-test}

\begin{algorithm}[tbp]
    \caption{TESLA safe packet test. For notations, see Table~\ref{tab:wire-format}.}
    \label{alg:safe-packet-test}
    \centering
    \procedure[linenumbering, codesize=\small]{$\mathsf{SafePacketTest}(t_{\mathrm{rx}},\; i,\; \Pi,\; D_t)$}{%
      \textbf{Output:}\; \top \text{ if safe, } \bot \text{ if unsafe} \pcskipln \\
      \bar{t}_{\mathrm{s}} \gets t_{\mathrm{rx}} + D_t \pccomment{upper bound on sender's time on rx} \\
      \bar{j} \gets \lfloor (\bar{t}_{\mathrm{s}} - T_0) \,/\, T_{\mathrm{int}} \rfloor \pccomment{upper bound on sender's interval} \\
      \pcif \bar{j} \geq i + d \pcthen \pcreturn \bot \pccomment{key $K_i$ may have been disclosed (unsafe)} \\
      \pcreturn \top
    }
    \end{algorithm}

Algorithm~\ref{alg:safe-packet-test} formalizes the safe packet test invoked by $\mathsf{TF5.VerifySIB1}$ (Algorithm~\ref{alg:ue-procedures}). Given a message of interval $i$ arriving at $t_{\mathrm{rx}}$, the receiver calculates the upper bound of the sender's current interval using $D_t$ and discards the message if the sender may already have reached the disclosure interval for $K_i$. Accepting such a message would allow an adversary who learned the disclosed $K_i$ to forge a valid MAC.

\subsection{Clock error bound calculation.}
\label{app:clock-error-bound}
Both the UE and the gNB calculate their clock error bounds at time $t$ using the following formula:
\begin{equation}
\label{eq:clock-error-bound}
\epsilon(t) = \epsilon_{\mathrm{sync}} + \rho\bigl(t-t_{\mathrm{sync}}\bigr) + \epsilon_{\mathrm{margin}},
\end{equation}
where $t_{\mathrm{sync}}$ is the last GPS synchronization time, $\epsilon_{\mathrm{sync}}$ is the initial error from the last synchronization, $\rho$ is the maximum clock drift rate, and $\epsilon_{\mathrm{margin}}$ is a safety margin.
$\epsilon_{\mathrm{sync}}$ is derived from the GPS receiver's timing accuracy specifications, and $\rho$ is derived from the oscillator's frequency error limits.
Both sets of specifications are already provided by component manufacturers in their datasheets~\cite{ubloxZEDF9P02B2024,sitimeSiT1532Datasheet2025,epsonTG2016SLA}.
\sysabbrv{} operates in the RRC layer and has to be implemented in the UE's modem firmware; therefore, UE manufacturers can incorporate these specifications into the clock error calculation during firmware implementation.

\section{Benchmark Scaling Methodology} \label{appendix:scaling}
\begin{table}[tbp]
\centering
\caption{\coremark{}/MHz~\cite{eembcCoreMark} and \ebacs{}~\cite{eBACS} cycle counts on reference (RPi 4B) and target (smartphone, V2X, and IoT) cores. The CM/MHz column shows the per-core \coremark{}/MHz score. SHA-256, Ed25519, and ECDSA columns show the \ebacs{} cycle counts for the respective primitives: SHA-256 on a 64-byte message; Ed25519 and ECDSA verify on a 59-byte message. Cortex-M33 is not in \ebacs{}; shown as {---}.}
\label{tab:scaling-platforms}
\begin{tabular}{llrrrr}
\hline
\textbf{Platform} & \textbf{Processor} & \textbf{CM/MHz} & \textbf{SHA-256} & \textbf{Ed25519 vrfy} & \textbf{ECDSA vrfy} \\
\hline
RPi 4B & Cortex-A72 & 5.668 & 1{,}539 & 406{,}698 & 641{,}007 \\
Smartphone & Cortex-A76 & 7.505 & 272     & 334{,}798 & 478{,}673 \\
V2X & Cortex-A53 & 3.201 & 2{,}195 & 603{,}383 & 953{,}118 \\
IoT & Cortex-M33 & 4.048 & ---     & ---       & ---       \\
\hline
\end{tabular}
\end{table}

To predict UE verification overhead for the smartphone, V2X, and IoT platforms in \S\ref{sec:eval:compute}, we scale per-operation computation time measured on the RPi~4B reference platform.
Table~\ref{tab:scaling-platforms} lists each core's \coremark{}/MHz score~\cite{eembcCoreMark} and \ebacs{}~\cite{eBACS} cycle counts.

\paragraph{Scaling Methodology.}
Let $T_{\text{ref}}$ be a measured computation time of the reference platform; we estimate the computation time of a target platform as:
\begin{equation}
\label{eq:scaling}
T_{\text{target}} \;=\; T_{\text{ref}} \cdot R \cdot \frac{f_{\text{ref}}}{f_{\text{target}}},
\end{equation}
where $f_{\text{ref}}$ and $f_{\text{target}}$ are the clock frequencies of the reference and target platforms, respectively, and the ratio $R$ takes one of two forms depending on the availability of \ebacs{} cycle counts for the two platforms.
When both processors appear in \ebacs{}, $R = C_{\text{target}}/C_{\text{ref}}$, where $C$ values are primitive-specific cycle counts reported by \ebacs{}.
Otherwise, $R$ falls back to $S_{\text{ref}}/S_{\text{target}}$, where $S$ values are per-core \coremark{}/MHz scores reported by \coremark{}. This fallback applies to BLS12-381 pairings, which are not measured in \ebacs{}, and to all IoT predictions because the Cortex-M33 is absent from \ebacs{}.
Values predicted with \ebacs{} carry \emph{high confidence}; Smartphone and V2X pairing predictions carry \emph{medium confidence}; all IoT predictions carry \emph{low confidence} due to the cross-ISA discontinuity between ARMv8-A and ARMv8-M (see Table~\ref{tab:eval-compute-results} for actual results).

\section{TESLA Evaluation in Prior Work.}
\label{app:prior-tesla-eval}

This section summarizes Hussain et al.'s~\cite{hussainInsecureConnectionBootstrapping2019} evaluation of TESLA in LTE SI authentication, which is the most relevant prior work to \sysabbrv{}.
Their experimental setup consists of separate LTE base station and UE equipment, each comprising an Intel Core i7 PC and a USRP B210, running a modified srsRAN LTE stack~\cite{srsRAN4G}.
They set $N = 10{,}000$ and employ the SCRA-BGLS signature scheme~\cite{yavuzRealTimeDigitalSignatures2017}.
Although they do not disclose the exact values of $T_{\mathrm{int}}$ and $d$, they configure the key disclosure delay, $d \cdot T_{\mathrm{int}}$, as 80\,ms.
In their experiment, acquiring SIB1 and SIB2 and verifying the signature takes 270\,ms; adding the 80\,ms key disclosure delay yields a total end-to-end delay of 350\,ms.
Of this, signature verification accounts for 110\,ms, while message verification using the MAC takes only 0.0028\,ms.
In terms of communication overhead, the bootstrap data (including the signature) occupies 220 bytes, the TESLA MAC 32 bytes, and the disclosed key 32 bytes, resulting in a total of 284 bytes.

Since their method rely on PKI, their bootstrap data is relatively large at 220 bytes, compared to 134 bytes in \sysabbrv{} (Table~\ref{tab:eval-comm-overhead}).
Moreover, SCRA-BGLS is a pairing-based, computationally heavy signature scheme, which results in a UE-side signature verification time of 110\,ms.
In contrast, \sysabbrv{} signature verification takes only ${\sim}0.042$\,ms on an Intel Core i5-14600KF at 5.3\,GHz (Table~\ref{tab:eval-compute-results}).\footnote{Hussain et al.~\cite{hussainInsecureConnectionBootstrapping2019} do not specify the exact Core i7 model used, which limits direct performance comparisons. Nevertheless, the reported verification times differ by more than three orders of magnitude with our measurements.}
A direct comparison of the end-to-end delays is difficult, however, because the key disclosure delays differ (80\,ms vs.\ 160\,ms) and the two evaluations use different RAN stacks.

\section{Tamarin Lemma Definitions}
\label{app:tamarin-lemmas}

This section presents only the two main lemmas discussed in paragraphs in \S\ref{sec:security}.
The full model linked there provides the model definitions and remaining lemmas.

In Tamarin, \texttt{All} and \texttt{Ex} denote universal and existential quantification.
The notation \texttt{E(...) @ i} records event \texttt{E} at trace position \texttt{i}, and \texttt{i < j} denotes event ordering.
In both lemmas, \texttt{nc\_comm} denotes the successor commitment $K_0^\mathrm{next}$, \texttt{m} the original SIB1 message, and \texttt{cid} the cell ID.

\medskip
\noindent\begin{minipage}{\linewidth}
\subsection{SIB1 authenticity}
\label{app:sib1-authenticity}

\texttt{SIB1\_Verified} records successful SIB1 authentication by the UE, and \texttt{SIB1\_Sent} records the gNB sending an SIB1 message.
\texttt{GnbCompromised} records the compromise of a gNB signing key.
\begin{small}
\begin{verbatim}
lemma sib1_authenticity:
  "All uid cid k0 t0 t_int d n m nc_comm new_authinput
       new_tag k_verify anchor t_rx #i.
    SIB1_Verified(
      uid,
      chain_ctx(cid, k0, t0, t_int, d, n),
      sib1_pkt(new_authinput, new_tag, k_verify),
      anchor,
      sib1_pkt(mac_authinput(cid, m, nc_comm),
        MAC(f_prime(k_verify), mac_authinput(cid, m, nc_comm)),
        f(k_verify)), t_rx
    ) @ i
    ==>
      (Ex #s.
        SIB1_Sent(chain_ctx(cid, k0, t0, t_int, d, n),
          sib1_pkt(mac_authinput(cid, m, nc_comm),
            MAC(f_prime(k_verify), mac_authinput(cid, m, nc_comm)),
            f(k_verify))) @ s
        & s < i)
    | (Ex exp #j.
        GnbCompromised(ibs_id(cid, exp)) @ j & j < i)"
\end{verbatim}
\end{small}
\end{minipage}

\newpage
\noindent\begin{minipage}{\linewidth}
\subsection{SIB1 freshness}
\label{app:sib1-freshness}

\texttt{SIB1\_Received} records SIB1 packet reception of the UE, and \texttt{IntervalExpired} records expiry of the interval associated with the key \texttt{k\_verify}.
\begin{small}
\begin{verbatim}
lemma sib1_freshness:
  "All uid cid k0 t0 t_int d n m nc_comm k_disclosed
       new_authinput new_tag k_verify anchor t_rx #r #v.
    SIB1_Received(
      uid,
      chain_ctx(cid, k0, t0, t_int, d, n),
      sib1_pkt(
        mac_authinput(cid, m, nc_comm),
        MAC(f_prime(k_verify), mac_authinput(cid, m, nc_comm)),
        k_disclosed), t_rx
    ) @ r
    & SIB1_Verified(
        uid,
        chain_ctx(cid, k0, t0, t_int, d, n),
        sib1_pkt(new_authinput, new_tag, k_verify),
        anchor,
        sib1_pkt(
          mac_authinput(cid, m, nc_comm),
          MAC(f_prime(k_verify), mac_authinput(cid, m, nc_comm)),
          k_disclosed), t_rx
      ) @ v
    & r < v
    ==>
      All #e.
        IntervalExpired(
          chain_ctx(cid, k0, t0, t_int, d, n), k_verify) @ e
      ==> r < e"
\end{verbatim}
\end{small}
\end{minipage}

\begin{credits}
\subsubsection{\ackname}
This work was supported by IITP under Next-generation Cloud-native Cellular Network Leadership Program (IITP-2026-RS-2024-00418784) grant funded by the Korea government (MSIT).
\end{credits}

\bibliographystyle{splncs04}
\bibliography{tf5-manual,tf5-zotero}

\end{document}